\documentclass[article,12pt,showpacs]{revtex4}
\usepackage{amsfonts,subfigure,graphicx,amsmath,mathtools,bm,multirow}

\begin{document}
\title[Short Title]{Robust entanglement between a movable mirror and atomic ensemble and entanglement transfer in coupled optomechanical system}
\author{Cheng-Hua Bai}
\affiliation{Department of Physics, College of Science, Yanbian
University, Yanji, Jilin 133002, People's Republic of China}
\author{Dong-Yang Wang}
\affiliation{Department of Physics, College of Science, Yanbian
University, Yanji, Jilin 133002, People's Republic of China}
\author{Hong-Fu Wang\footnote{E-mail: hfwang@ybu.edu.cn}}
\affiliation{Department of Physics, College of Science, Yanbian
University, Yanji, Jilin 133002, People's Republic of China}
\author{Ai-Dong Zhu}
\affiliation{Department of
Physics, College of Science, Yanbian University, Yanji, Jilin
133002, People's Republic of China}
\author{Shou Zhang\footnote{E-mail: szhang@ybu.edu.cn}}
\affiliation{Department of
Physics, College of Science, Yanbian University, Yanji, Jilin
133002, People's Republic of China}

\begin{abstract}
We propose a scheme for the creation of robust entanglement between a movable mirror and atomic ensemble at the macroscopic level in coupled optomechanical system. In experimentally accessible parameter regimes, we show that critical temperature of the bipartite continuous variable entanglement in our scheme can be raised from previous 24 K [Vitali {\it et al.}, Phys. Rev. Lett. \textbf{98}, 030405 (2007)] and 20 K [Genes {\it et al.}, Phys. Rev. A \textbf{77}, 050307(R) (2008)] to 32 K. We also investigate the entanglement transfer based on this coupled system. The scheme can be used for the realization of quantum memories for continuous variable quantum information processing and quantum-limited displacement measurements.
\end{abstract}
\maketitle

\section{Introduction}\label{sec1}
Entanglement, the characteristic trait of quantum mechanics~\cite{ESPCPSOC3531}, has appealed widespread attention and interest in different branches of physics and become the essential resource for many quantum information processes~\cite{DBAEAZ2000,MANILC2000}. Up to now, entanglement has been successfully prepared and manipulated in variously microscopic system theoretically and experimentally, such as atoms~\cite{HASKNJP1113,WMSSLSZJYLADZHFWSZJOSAB1532,SLSXQSHFWSZPRA1490,SLSXQSHFWSZSR144,SLSQGHFWSZPRA1592}, photons\cite{HFWSZPRA0979,HFWSZEPJD0953,HSAXKOE1119}, ions~\cite{JPPRL9574,DCDN02417,ACWYCKRBEKDLDJWNature14512}, and so on. However, it is not yet completely clear that to what degree quantum mechanics is suitable for mesoscopic and macroscopic systems~\cite{DVSGAFHRBPTAGVVAZMAPRL0798}. In addition, the preparation of entanglement in mesoscopic and macroscopic systems should not be violated in principle~\cite{DVSGAFHRBPTAGVVAZMAPRL0798,YHMLZJAP12111}, but quantum phenomena such as quantum entanglement commonly does not present in the macroscopic world resulting form the environment-induced decoherence. So many scientists tried their best to observe the novel phenomena of obtaining quantum entanglement at the mesoscopic and macroscopic level over the last decades. Fortunately, due to the theoretical investigation and the advancement of experimental techniques, obtaining and observing entanglement in mesocopic and even macroscopic systems have become possible.

With the fast-developing field of microfabrication and nanotechnology, cavity optomechanical system is being one of the most appealing and promising candidates as an ideal system for the study of fundamental quantum physics, such as macroscopic quantum phenomena, decoherence, and quantum-classical boundary~\cite{YCLYWHCWWYFXCPB1322}. Its standard model consists of a Fabry-P\'{e}rot cavity with a fixed partially transmitting mirror and one movable perfectly reflecting mirror. When the Fabry-P\'{e}rot cavity is coherently driven by an external laser field, the movable mirror will be shifted from its equilibrium position and free to move along the cavity axis due to the radiation pressure force and its center-of-mass motion can be modelled as a mechanical harmonic oscillator. Since the mechanical harmonic oscillator resembles a prototype of classical systems~\cite{DVSGAFHRBPTAGVVAZMAPRL0798}, cavity optomechanical system provides a unique platform for exploring the novel quantum phenomena at the mesoscopic and macroscopic level such as optomechanical entanglement. In recent years, a number of schemes covering this topic have been proposed based on the cavity optomechanical system~\cite{DVSGAFHRBPTAGVVAZMAPRL0798,CGDVPTPRA0877,CGAMPTDVPRA0878,CJJLMJEAPOPRA1285,UAWMKNGJMPRA1286,WGMAAHNMSZPRA1388,JQLQQWFNPRA1489,THRZHIPRA1592,QWYXZMZCPB1524}. In 2007, Vitali {\it et al.} successfully generated stationary entanglement between an optical cavity field mode and a macroscopic vibrating mirror in a standard optomechanical setup and showed that such optomechanical entanglement was robust against the environment temperature above 20 K~\cite{DVSGAFHRBPTAGVVAZMAPRL0798}. Latter, their group also realized the tripartite and bipartite continuous variable entanglement by placing an ensemble of two-level atoms inside the Fabry-P\'{e}rot cavity~\cite{CGDVPTPRA0877}. Then they realized the optomechanical entanglement between the experimentally detectable output field of an optical cavity and a vibrating cavity end-mirror~\cite{CGAMPTDVPRA0878}. In 2012, Joshi {\it et al.} investigated the possibility of generating optomechanical entanglement between optical and mechanical modes of two spatially separated cavities theoretically in which each cavity was assumed to have one fixed and one movable mirror and the two cavities were coupled by an optical fiber~\cite{CJJLMJEAPOPRA1285}. The entanglement between the different optical and mechanical modes in an array of three coupled optomechanical cavities was considered and the dynamics of such a setup showed that intracavity optomechanical entanglement generated independently in each cavity can be distributed pairwise between intercavity photons as well as phonons~\cite{UAWMKNGJMPRA1286}. Subsequently, the scheme for entangling two macroscopic mechanical resonators (movable mirrors) by their coupling to the two-mode fields of a correlated-emission laser inside a doubly resonant cavity was proposed~\cite{WGMAAHNMSZPRA1388}, showing that the steady-state entanglement of two mirrors as well as that of two-mode fields can be obtained in the regime of strong field-mirror coupling when the input lasers are scattered at the anti-Stokes sidebands. In 2014, Liao {\it et al.} proposed a scheme to generate quantum entanglement between two macroscopic mechanical resonators in a two-cavity optomechanical system~\cite{JQLQQWFNPRA1489}. In a double-cavity system of a mechanical resonator coupled to two cavity modes on both sides through radiation pressure, Huan {\it et al.} investigated entanglement transfer from the intracavity photon-phonon entanglement to an intercavity photon-photon entanglement~\cite{THRZHIPRA1592}. Wu {\it et al.} investigated the entanglement properties in a hybrid system consisting of an optical cavity-array coupled to a mechanical resonator in 2015~\cite{QWYXZMZCPB1524}. Here we propose a scheme for the creation of robust entanglement between a movable mirror and atomic ensemble at the macroscopic level in coupled optomechanical system. In the scheme, with the increase of the coupling strength of the coupled optomechanical system, not only the entanglement is increasingly stronger but also the effective detuning is more and more broader, which are extremely significant due to the fact that the stronger entanglement and the more broader effective detuning are obtained, the more easily it is realized and observed in experiment. In the meanwhile, the numerical simulation results indicate that critical temperature of the bipartite continuous variable entanglement can be raised from 24 K in Ref.~\cite{DVSGAFHRBPTAGVVAZMAPRL0798} and 20 K in Ref.~\cite{CGDVPTPRA0877} to 32 K in our scheme in experimentally accessible parameter regimes. Moreover, we also investigate the entanglement transfer based on this coupled system. Our scheme can be used for the realization of quantum memories for continuous variable quantum information processing and quantum-limited displacement measurements.

The remainder of this paper is organized as follows. In Sec. \uppercase\expandafter{\romannumeral 2} we establish the theoretical model of the coupled optomechanical system and present the equations of motion of the system. In Sec. \uppercase\expandafter{\romannumeral 3} we quantify the entanglement properties of the system by introducing the logarithmic negativity. In an experimentally accessible parameter regime, we simulate the entanglement properties of the coupled optomechanical system numerically in Sec. \uppercase\expandafter{\romannumeral 4}. Finally we make a conclusion to summarize our results in Sec. \uppercase\expandafter{\romannumeral 5}.

\section{The coupled optomechanical system model and equations of motion}\label{sec2}
As schematically shown in Fig.~1, the system studied here is composed of two coupled single-mode cavities and an ensemble of two-level atoms. Cavity 1 contains the atomic ensemble and is coherently driven by an external monochromatic laser field with strength $\Omega_l$ and frequency $\omega_l$ and cavity 2 with a fixed mirror and a second oscillating mirror couples to the cavity 1 with the coupling strength $J$. The optical field of cavity 2 is coupled to the mechanical motion of the movable mirror via radiation pressure force and mirror vibrational motion can be modelled as a mechanical harmonic oscillator of frequency $\omega_m$ and decay rate $\gamma_m$. Experimentally, such a double-cavity optomechanical model can be carried out in the systems based on Fabry-P\'{e}rot cavities or whispering-gallery cavities~\cite{ISGHLOPKJVPRL10104,BPSKOFLFMMGGLLSFFNCMBLYNP1410,LCXJSHCYJWLJGLGWMXNP148}. The Hamiltonian for describing the coupled optomechanical system is written as~\cite{CKLPRA9551}
\begin{eqnarray}\label{e1}
H&=&\sum_{j=1}^{2}\hbar\omega_ja_j^\dag a_j+\frac{\hbar}{2}\omega_a S_z+\frac{\hbar}{2}\omega_m(q^2+p^2)+\hbar J(a_1^\dag a_2+a_1a_2^\dag)\cr\cr
&&+\hbar g(S_+a_1+S_-a_1^\dag)-\hbar G_0a_2^\dag a_2q+i\hbar\Omega_l(a_1^\dag e^{-i\omega_l t}-a_1e^{i\omega_l t}),
\end{eqnarray}
where $a_j$ is the bosonic operator eliminating a photon in the $j$-th cavity with resonance frequency $\omega_j$. The atomic ensemble is composed of $N$ two-level atoms with intrinsic frequency $\omega_a$ each described by the spin-1/2 Pauli matrices $\sigma_+$, $\sigma_-$, and $\sigma_z$. Collective spin operators are defined as $S_{+,-,z}=\sum\limits_{i=1}^{N}\sigma_{+,-,z}^{(i)}$ and satisfy the commutation relations $[S_+,S_-]=S_z$ and $[S_z,S_{\pm}]=\pm2S_{\pm}$. $q$ and $p$ are the dimensionless position and momentum operators of the oscillating mirror, respectively, and satisfy $[q, p]=i$. $g$ is the atom-cavity coupling constant and given by $g=\mu\sqrt{\omega_1/2\hbar\epsilon_{0}V}$, where $\mu$ is the dipole moment of the atomic transition, $\epsilon_0$ is the free space permittivity, and $V$ is the volume of cavity 1 mode. $G_0=(\omega_2/L)\sqrt{\hbar/m\omega_m}$ is the radiation pressure coupling strength, with $L$ the cavity length in the absence of the intracavity field and $m$ the effective mass of the mechanical mode~\cite{MPYHAHEPJD997}. The strong drive of amplitude $\Omega_l=\sqrt{2P\kappa/\hbar\omega_l}$, with $P$ and $\kappa$ the drive laser input power and the cavity decay rate, respectively, resulting in a large steady-state optical field in the cavity, which increases the occupation numbers in each mode and the radiation pressure coupling. The induced steady-state intracavity in turn shifts the equilibrium position of the mechanical oscillator via the radiation pressure force. In Eq.~({\ref{e1}}), the first three terms denote the free energy of the coupled optomechanical system, the fourth term represents the coupling between the cavity 1 and cavity 2, the fifth term describes the coupling of atomic ensemble with cavity mode, the sixth term represents the coupling of optical mode with mechanical mode, and the last term describes the coupling of laser drive with the cavity, respectively.
\begin{figure}\label{fig1}
\centering
\includegraphics[width=4.5in,height=2.78in]{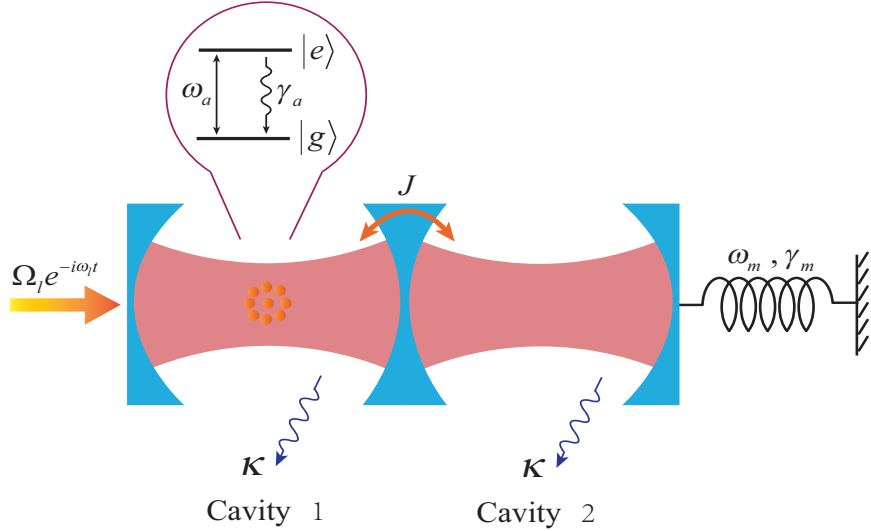}
\caption{(Color online) Schematic illustration of the coupled optomechanical system including cavity 1 coupled to cavity 2 with coupling strength $J$. An ensemble of two-level atoms is placed into the cavity 1 which is coherently driven by an external monochromatic laser field with strength $\Omega_l$ and frequency $\omega_l$. The vibrational motion of the oscillating mirror for cavity 2 can be modelled as a mechanical harmonic oscillator of frequency $\omega_m$ and decay rate $\gamma_m$ and is shifted from the equilibrium position due to the radiation pressure force.}
\end{figure}

We consider a compact scenario of such a equation, which is accessible in the low atomic excitation limit, i.e., the average number of atoms in the excited state $|e\rangle$ is much smaller than the number of total atoms~\cite{SBZPRA1286}. In this limit, the collective spin operators ${S_{\pm}, S_z}$ of the atomic polarization can be described in terms of the bosonic annihilation and creation operators $c$ and $c^\dag$ via the Holstein-Primakoff transformation~\cite{THHPPR4058,KHASSESPRMP1082}
\begin{eqnarray}\label{e2}
S_+&=&c^\dag\sqrt{N-c^\dag c}\simeq\sqrt{N}c^\dag,\cr\cr
S_-&=&\sqrt{N-c^\dag c}c\simeq\sqrt{N}c,\cr\cr
S_z&=&c^\dag c-N/2\simeq -N/2,
\end{eqnarray}
where the usual bosonic commutation relation $[c, c^\dag]=1$ is still satisfied.

Transforming the above Hamiltonian into the rotating frame at the frequency $\omega_l$ of the driving laser field, we rewrite the system Hamiltonian as
\begin{eqnarray}\label{e3}
H^\prime&=&\sum_{j=1}^{2}\hbar\Delta_j a_{j}^{\dag}a_j+\hbar\Delta_ac^{\dag}c+\frac{\hbar}{2}\omega_m(q^2+p^2)+\hbar J(a_1^{\dag}a_2+a_1a_2^{\dag})\cr\cr
&&+\hbar G_a(c^{\dag}a_1+c a_1^{\dag})-\hbar G_0a_2^{\dag}a_2q+i\hbar \Omega_l(a_1^{\dag}-a_1),
\end{eqnarray}
where $\Delta_j=\omega_j-\omega_l$ and $\Delta_a=\omega_a-\omega_l$ are, respectively, the cavity mode and atomic detuning with respect to the driving laser, $G_a=g\sqrt{N}$.

A proper analysis of the dynamics of the coupled optomechanical system can be accomplished by a set of nonlinear Langevin equations in which the corresponding dissipation and fluctuation terms are added to the Heisenberg equations of motion derived form the Eq.~({\ref{e3}})
\begin{eqnarray}\label{e4}
\dot{q}&=&\omega_mp,\cr\cr
\dot{p}&=&-\omega_mq+G_0a_2^{\dag}a_2-\gamma_mp+\xi,\cr\cr
\dot{a_1}&=&-(\kappa+i\Delta_1)a_1-iG_ac+\Omega_l-iJa_2+\sqrt{2\kappa}a_1^{in},\cr\cr
\dot{a_2}&=&-(\kappa+i\Delta_2)a_2-iJa_1+iG_0a_2q+\sqrt{2\kappa}a_2^{in},\cr\cr
\dot{c}&=&-(\gamma_a+i\Delta_a)c-iG_aa_1+\sqrt{2\gamma_a}c^{in},
\end{eqnarray}
where $\gamma_a$ is the decay rate of the atomic excited state level and the nonvanishing correlation functions of noises affecting atoms and cavity fields obey the relations $\langle a_j^{in}(t)a_j^{\dag in}(t^{\prime})\rangle=\langle c^{in}(t)c^{\dag in}(t^{\prime})\rangle=\delta(t-t^{\prime})$~\cite{DFWGJMQO1994,CWGPZQN2000}. Here we have assumed that the cavity 1 and cavity 2 have the same decay rate $\kappa$. Furthermore, the mechanical mode is also affected by the stochastic Hermitian Brownian noise $\xi$ that satisfies the non-Markovian correlation function with a colored spectrum in general~\cite{CWGPZQN2000}
\begin{eqnarray}\label{e5}
\langle\xi(t)\xi(t^{\prime})\rangle=\frac{\gamma_m}{\omega_m}\int\frac{\omega}{2\pi}e^{-i\omega(t-t^{\prime})}\left[\coth\left(\frac{\hbar\omega}{2k_BT}\right)+1\right]d\omega,
\end{eqnarray}
where $k_B$ is the Boltzmann constant and $T$ is the temperature of the mechanical oscillator. However, quantum effects are revealed only for the mechanical oscillator with a high quality factor, i.e., $Q=\omega_m/\gamma_m\gg1$. In this limit, this non-Markovian process can be approximated as a Markovian one and the Brownian noise $\xi(t)$ can be simplified to delta-correlated~\cite{RBMKPRL8146,VGDV0163}
\begin{eqnarray}\label{e6}
\langle\xi(t)\xi(t^{\prime})+\xi(t^{\prime})\xi(t)\rangle/2\simeq\gamma_m(2\bar{n}+1)\delta(t-t^{\prime}),
\end{eqnarray}
where $\bar{n}=(\mathrm{exp}\{\hbar\omega_m/k_BT\}-1)^{-1}$ is the mean thermal excitation number. In the following we discuss the entanglement of the coupled optomechanical system in the regime where the system is stable.
\section{The steady-state entanglement of the coupled optomechanical system}\label{sec3}
We now begin to linearize the dynamics of the coupled optomechanical system. The nonlinear quantum Langevin equations can be linearized by rewriting each Heisenberg operator as a sum of its steady-state mean value and an additional fluctuation operator with zero-mean value, i.e., $q=q_s+\delta q$, $p=p_s+\delta p$, $a_j=a_{js}+\delta a_j~(j=1, 2)$, and $c=c_s+\delta c$~\cite{MATJKFMRMP1486}. Substituting these expressions into Eq.~({\ref{e4}}) and the latter will be separated into a set of nonlinear algebra equations for the steady-state value and a set of quantum Langevin equations for the fluctuation operators~\cite{CFMPSBAHEGST9449}. The steady-state mean values of the coupled optomechanical system can be obtained by setting the time derivatives to zero,
\begin{eqnarray}\label{e7}
p_s&=&0,\cr\cr
q_s&=&G_0|a_{2s}|^2/\omega_m,\cr\cr
a_{2s}&=&-iJa_{1s}/(\kappa+i\Delta_2^{\prime}),\cr\cr
a_{1s}&=&\Omega_l/\left[\kappa+i\Delta_1+G_a^2/(\gamma_a+i\Delta_a)+J^2/(\kappa+i\Delta_2^{\prime})\right],\cr\cr
c_s&=&-iG_aa_{1s}/(\gamma_a+i\Delta_a),
\end{eqnarray}
where $\Delta_2^{\prime}=\Delta_2-G_0q_s$ is the effective detuning of the second cavity mode.

We assume that the cavity is intensively driven with a very large input power $P$, so that at the steady state, the intracavity fields have a large amplitude $a_{js}$, i.e., $|a_{js}|\gg1$. In the strong driving limit, we can safely omit the nonlinear quantities $\delta a_2^{\dag}\delta a_2$ and $\delta a_2\delta q$ and get the following linearized Langevin equations,
\begin{eqnarray}\label{e8}
\delta\dot{q}&=&\omega_m\delta p,\cr\cr
\delta\dot{p}&=&-\omega_m\delta q+G_0a_{2s}(\delta a_2+\delta a_2^{\dag})-\gamma_m\delta p+\xi,\cr\cr
\delta\dot{a_1}&=&-(\kappa+i\Delta_1)\delta a_1-iG_a\delta c-iJ\delta a_2+\sqrt{2\kappa}a_1^{in},\cr\cr
\delta\dot{a_2}&=&-(\kappa+i\Delta_2)\delta a_2-iJ\delta a_1+iG_0(a_{2s}\delta q+q_s\delta a_2)+\sqrt{2\kappa}a_2^{in},\cr\cr
\delta\dot{c}&=&-(\gamma_a+i\Delta_a)\delta c-iG_a\delta a_1+\sqrt{2\gamma_a}c^{in},
\end{eqnarray}
where we have chosen the phase reference of the cavity fields so that $a_{js}$ can be taken real. Here we will devote to establishing the presence of quantum correlations among the subsystems of the coupled optomechanical system at the steady state, which can be carried out by analyzing the dynamics of the quantum fluctuations of the coupled optomechanical system around the steady state. Generally, it is convenient to use the pairwise quadrature operators defined by
\begin{eqnarray}\label{e9}
\left\{\begin{lgathered}
\delta X_1=(\delta a_1+\delta a_1^{\dag})/\sqrt{2}\\
\delta Y_1=(\delta a_1-\delta a_1^{\dag})/i\sqrt{2}
\end{lgathered}\right.,
\end{eqnarray}
\begin{eqnarray}\label{e10}
\left\{\begin{lgathered}
\delta X_2=(\delta a_2+\delta a_2^{\dag})/\sqrt{2}\\
\delta Y_2=(\delta a_2-\delta a_2^{\dag})/i\sqrt{2}
\end{lgathered}\right.,
\end{eqnarray}
\begin{eqnarray}\label{e11}
\left\{\begin{lgathered}
\delta x=(\delta c+\delta c^{\dag})/\sqrt{2}\\
\delta y=(\delta c-\delta c^{\dag})/i\sqrt{2}
\end{lgathered}\right.,
\end{eqnarray}
and the corresponding Hermitian input noise operators
\begin{eqnarray}\label{e12}
\left\{\begin{lgathered}
X_1^{in}=(a_1^{in}+a_1^{in\dag})/\sqrt{2}\\
Y_1^{in}=(a_1^{in}-a_1^{in\dag})/i\sqrt{2}
\end{lgathered}\right.,
\end{eqnarray}
\begin{eqnarray}\label{e13}
\left\{\begin{lgathered}
X_2^{in}=(a_2^{in}+a_2^{in\dag})/\sqrt{2}\\
Y_2^{in}=(a_2^{in}-a_2^{in\dag})/i\sqrt{2}
\end{lgathered}\right.,
\end{eqnarray}
\begin{eqnarray}\label{e14}
\left\{\begin{lgathered}
x^{in}=(c^{in}+c^{in\dag})/\sqrt{2}\\
y^{in}=(c^{in}-c^{in\dag})/i\sqrt{2}
\end{lgathered}\right..
\end{eqnarray}
Then the resulting evolution equations of motion for the fluctuations in Eq.~({\ref{e8}}) can be rewritten in a compact form as follows,
\begin{eqnarray}\label{e15}
\dot{\textbf{u}}=\textbf{A}\textbf{u}+\textbf{n},
\end{eqnarray}
where the vector of eight-component quadrature fluctuations $\textbf{u}=(\delta q, \delta p, \delta X_1, \delta Y_1, \delta X_2, \delta Y_2,\\ \delta x, \delta y)^T$, similarly the input-noise vector $\textbf{n}=(0, \xi, \!\!\sqrt{2\kappa}X_1^{in}, \!\!\sqrt{2\kappa}Y_1^{in}, \!\!\sqrt{2\kappa}X_2^{in}, \!\!\sqrt{2\kappa}Y_2^{in},\!\! \sqrt{2\gamma_a}x^{in}\\, \sqrt{2\gamma_a}y^{in})^T$, and the drift matrix $\textbf{A}$ is given by
\begin{eqnarray}\label{e16}
\textbf{A}=
\left(
\begin{array}{cccccccc}
0~&\omega_m~&0~&0~&0~&0~&0~&0\\
-\omega_m~&-\gamma_m~&0~&0~&G~&0~&0~&0\\
0~&0~&-\kappa~&\Delta_1~&0~&J~&0~&G_a\\
0~&0~&-\Delta_1~&-\kappa~&-J~&0~&-G_a~&0\\
0~&0~&0~&J~&-\kappa~&\Delta_2^{\prime}~&0~&0\\
G~&0~&-J~&0~&-\Delta_2^{\prime}~&-\kappa&0~&0\\
0~&0~&0~&G_a~&0~&0~&-\gamma_a~&\Delta_a\\
0~&0~&-G_a~&0~&0~&0~&-\Delta_a~&-\gamma_a
\end{array}
\right),
\end{eqnarray}
where $G=\sqrt{2}G_0a_{2s}$ is the effective optomechanical coupling. Now the quantum fluctuations of the field and the oscillating mirror are coupled by the much larger effective optomechanical coupling $G$, so the engineering of significant optomechanical entanglement in coupled system becomes possible~\cite{DVSGAFHRBPTAGVVAZMAPRL0798}.

The formal solution of the first-order linear inhomogeneous differential Equation~({\ref{e15}}) can be expressed as
\begin{eqnarray}\label{e17}
\textbf{u}(t)=\textbf{f}(t)\textbf{u}(0)+\int_0^t\textbf{f}(\tau)\textbf{n}(t-\tau)d\tau,
\end{eqnarray}
where the matrix $\textbf{f}(t)=\mathrm{exp}\{\textbf{A}t\}$ and the initial condition $\textbf{f}(0)=\textbf{{\it I}}$ ($\textbf{{\it I}}$ is the identity matrix). In the present coupled system, appealing quantities are the quadrature fluctuations. Thus we define a covariance matrix $\textbf{V}(t)$ by the element $\textbf{V}_{ij}(t)=\frac12(\langle\textbf{u}_i(t)\textbf{u}_j(t)+\textbf{u}_j(t)\textbf{u}_i(t)\rangle)$ for $i,j=1,2,\cdots,8$. The coupled system is stable and reaches its steady state only if the real part of all the eigenvalues of the drift matrix $\textbf{A}$ are negative so that $\textbf{f}(\infty)=0$. The stability conditions can be derived by applying the Routh-Hurwitz criterion~\cite{EXDKPRA8735} and the case of eight dimensions matrix is shown in Ref.~\cite{WGMAAHNMSZPRA1388}. We will guarantee the stability conditions of the system in the following analysis. When the coupled system reaches its steady state ($t\rightarrow\infty$), one can obtain
\begin{eqnarray}\label{e18}
\textbf{V}_{ij}=\sum_{k,l}\int_0^{\infty}d\tau\int_0^{\infty}d\tau^{\prime}\textbf{f}_{ik}(\tau)\textbf{f}_{jl}(\tau^{\prime})
\bm{\Phi}_{kl}(\tau-\tau^{\prime}),
\end{eqnarray}
where $\bm{\Phi}_{kl}(\tau-\tau^{\prime})=\frac12(\langle\textbf{n}_k(\tau)\textbf{n}_l(\tau^{\prime})+\textbf{n}_l(\tau^{\prime})\textbf{n}_k(\tau)\rangle)$ is the stationary noise correlation matrix. Due to the fact that the seven components of $\textbf{n}(t)$ are uncorrelated, we can get $\bm{\Phi}_{kl}(\tau-\tau^{\prime})=\textbf{D}_{kl}\delta(\tau-\tau^{\prime})$, where $\textbf{D}=\mathrm{Diag}[0, \gamma_m(2\bar{n}+1), \kappa, \kappa, \kappa, \kappa, \gamma_a, \gamma_a]$ is the diagonal matrix for the corresponding damping and leakage rates stemming from the noise correlations. Hence Eq.~({\ref{e18}}) becomes $\textbf{V}=\int_0^{\infty}\textbf{f}(\tau)\textbf{D}\textbf{f}(\tau)^Td\tau$. When the stability conditions of the coupled system are satisfied, the steady-state correlation matrix can be derived from the following Lyapunov equation~\cite{DVSGAFHRBPTAGVVAZMAPRL0798,PCPVHST1993,MBPMPRL0799}
\begin{eqnarray}\label{e19}
\textbf{A}\textbf{V}+\textbf{V}\textbf{A}^T=-\textbf{D}.
\end{eqnarray}
From this equation, the covariance matrix $\textbf{V}$ can be written as the form of a block matrix
\begin{eqnarray}\label{e20}
\textbf{V}=
\left(
\begin{array}{cccc}
\textbf{V}_m&\textbf{V}_{mc_1}&\textbf{V}_{mc_2}&\textbf{V}_{ma}\\
\textbf{V}_{mc_1}^T&\textbf{V}_{c_1}&\textbf{V}_{c_1c_2}&\textbf{V}_{c_1a}\\
\textbf{V}_{mc_2}^T&\textbf{V}_{c_1c_2}^T&\textbf{V}_{c_2}&\textbf{V}_{c_2a}\\
\textbf{V}_{ma}^T&\textbf{V}_{c_1a}^T&\textbf{V}_{c_2a}^T&\textbf{V}_{a}
\end{array}
\right),
\end{eqnarray}
where each block represents $2\times2$ matrix. The blocks on the diagonal indicate the variance within each subsystem (the oscillating mirror, the cavity mode 1, the cavity mode 2, and the atomic ensemble), while the off-diagonal blocks indicate covariance across different subsystems, i.e., the correlations between two components of the whole coupled optomechanical system.

To compute the entanglement among the subsystems of the coupled optomechanical system, we reduce the $8\times8$ covariance matrix \textbf{V} to a $4\times4$ submatrix $\textbf{V}_S$. If the indices $i$ and $j$ for the element $\textbf{V}_{ij}$ are confined to the set $\{1, 2, 3, 4\}$, the submatrix $\textbf{V}_S=[\textbf{V}_{ij}]$ is formed by the first four rows and columns of $\textbf{V}$ and corresponds to the covariance between the cavity 1 mode and the oscillating mirror. Similarly, if the indices run over $\{1, 2, 5, 6\}$, $\textbf{V}_S$ is the covariance matrix of the cavity mode 2 and the oscillating mirror. If the indices run over $\{1, 2, 7, 8\}$, $\textbf{V}_S$ labels the covariance between the atomic ensemble and the oscillating mirror. Summarizing, the submatrix can be written as
\begin{eqnarray}\label{e21}
\textbf{V}_S=
\left(
\begin{array}{cc}
\textbf{V}_m&\textbf{V}_{m\beta}\\
\textbf{V}_{m\beta}^{T}&\textbf{V}_\beta
\end{array}
\right),
\end{eqnarray}
where $m$ and $\beta$ index the subsystem \{oscillating mirror, cavity 1 (cavity 2, atomic ensemble)\} in the coupled optomechanical system.

Next we resort to Simon's criterion to judge continuous variable entanglement~\cite{RSPRL0084}. For a physical state, the covariance matrix $\textbf{V}$ must obey the Robertson-Schr\"{o}dinger uncertainty principle
\begin{eqnarray}\label{e22}
\textbf{V}+\frac{i}{2}\bm{\beta}\geq0,
\end{eqnarray}
where $\bm{\beta}=
\left(
\begin{array}{cc}
\textbf{J}&0\\
0&\textbf{J}
\end{array}
\right)$
with $\textbf{J}=
\left(
\begin{array}{cc}
0&1\\
-1&0
\end{array}
\right)$.
Here we define the vector $\textbf{F}=(Q_1, P_1, Q_2, P_2)^T$ for a two-mode system. If a state is separable, partial transpose matrix $\widetilde{\textbf{V}}$ (obtained from $\textbf{V}$ just by taking $P_j$ in $-P_j$) still comply with the Eq.~({\ref{e22}}). This inequality equation requires that all the symplectic eigenvalues of the transposed matrix to be larger than 1/2. So if the smallest eigenvalues is less than 1/2, the transposed modes are inseparable, i.e., there exists entanglement between the modes. We introduce the logarithmic negativity to quantify the entanglement which can be computed by means of a process known as symplectic diagonalization of submatrix $\textbf{V}_S$, where the entanglement properties are characterized in the symplectic eigenvalues of the diagonalized matrix. If the diagonalized matrix is written as $\mathrm{Diag}[\nu_-, \nu_-, \nu_+, \nu_+]$, then the eigenvalues along the diagonal is~\cite{MPSVQIC077}
\begin{eqnarray}\label{e23}
\nu_{\mp}=\sqrt{\tfrac12[\Sigma(\textbf{V}_S)\mp\sqrt{\Sigma(\textbf{V}_S)^2-4\mathrm{det}\textbf{V}_S}]},
\end{eqnarray}
where $\Sigma(\textbf{V}_S)=\mathrm{det}\textbf{V}_m+\mathrm{det}\textbf{V}_\beta-2\mathrm{det}\textbf{V}_{m\beta}$. We regard $\nu_-$ as the minimum sysplectic eigenvalue of the covariance matrix and the logarithmic negativity $E_N$ can be defined as $E_N=\mathrm{max}[0, -\mathrm{ln}2\nu_-]$~\cite{GAASFIPRA0470}. Therefore, the symplectic eigenvalue $\nu_-$ completely quantifies the quantum entanglement among the subsystems and they are entangled if and only if $\nu_-<\frac12$, which is consistent with the Simon's criterion. In the next section, we utilize the logarithmic negativity $E_N$ to show the entanglement properties of the coupled optomechanical system numerically.
\section{Numerical results of the entanglement among the subsystems and discussion}\label{sec4}
\begin{table}\label{1}
\caption{The experimental parameters for the coupled optomechanical system used in our numerical simulation, extracted from the experiments in Refs.~\cite{DKWMMJADKNDBJPWTMIDBPRL0696,SGHRBMPFBGLJBHKCSDBMAAZNature06444,OAPFCTBMPAHNature06444}.}
\begin{tabular}{ccc}\hline\hline
~~~~~~~~~Systematic parameter~~~~~~~~~&~~~~~~~~~Symbol~~~~~~~~~&~~~~~~~~~Value~~~~~~~~~~\\\hline
~~~~~~~~~Cavity length~~~~~~~~~&~~~~~~~~~$L$~~~~~~~~~&~~~~~~~~~1 mm~~~~~~~~~\\
~~~~~~~~~Cavity decay rate~~~~~~~~~&~~~~~~~~~$\kappa$~~~~~~~~~&~~~~~~~~~$\pi\times10^7$ Hz~~~~~~~~~\\
~~~~~~~~~Driven-laser wavelength~~~~~~~~~&~~~~~~~~~$\lambda$~~~~~~~~~&~~~~~~~~~810 nm~~~~~~~~~\\
~~~~~~~~~Input laser power~~~~~~~~~&~~~~~~~~~$P$~~~~~~~~~&~~~~~~~~~35 mW~~~~~~~~~\\
~~~~~~~~~Mechanical frequency~~~~~~~~~&~~~~~~~~~$\omega_m$~~~~~~~~~&~~~~~~~~~$2\pi\times10^7$ Hz~~~~~~~~~\\
~~~~~~~~~Mechanical mass~~~~~~~~~&~~~~~~~~~$m$~~~~~~~~~&~~~~~~~~~5 ng~~~~~~~~~\\
~~~~~~~~Mechanical damping rate~~~~~~~~&~~~~~~~~$\gamma_m$~~~~~~~~&~~~~~~~~$200\pi$ Hz~~~~~~~~\\
~~~~~~~~~Atomic decay rate~~~~~~~~~&~~~~~~~~~$\gamma_a$~~~~~~~~~&~~~~~~~~~$\pi\times10^7$ Hz~~~~~~~~~\\
~~~~~~~~~Atomic ensemble coupling strength~~~~~~~~~&~~~~~~~~~$G_a$~~~~~~~~~&~~~~~~~~~$1.2\pi\times10^7$ Hz~~~~~~~~~\\\hline\hline
\end{tabular}
\end{table}

In the following, we investigate the stationary optomechanical entanglement among the subsystems numerically. In fact, Eq.~({\ref{e19}}) is a linear equation for \textbf{V} and can be straightforwardly solved, but the general exact expression is too tedious. However, it is easy to simulate numerically. In our numerical calculations, we adopt the set of experimental parameters for the coupled optomechanical system given in Table \uppercase\expandafter{\romannumeral 1}, which can be carried out in current experiments~\cite{DKWMMJADKNDBJPWTMIDBPRL0696,SGHRBMPFBGLJBHKCSDBMAAZNature06444,OAPFCTBMPAHNature06444}, consequently, our scheme is experimentally feasible. In order to produce continuous variable entanglement in coupled optomechanical system, we must construct the effective Hamiltonian of the nondegenerate parametric-down conversion type for the system through setting $\Delta_1=-\Delta_2^{\prime}=-\Delta$~\cite{TYWZLZ1461}.
\begin{figure}\label{fig2}
\centering
\includegraphics[scale=0.6]{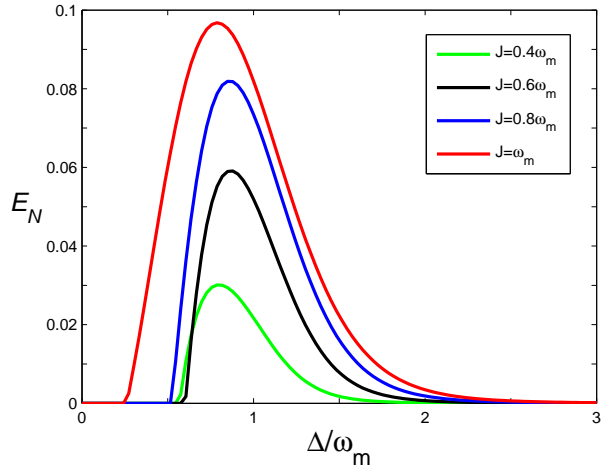}
\caption{(Color online) The logarithmic negativity $E_N$ as a function of the normalized detuning $\Delta/\omega_m$ with four different optical coupling strength $J$ at a fixed temperature $T$=400 mK and the other parameters are given in Table \uppercase\expandafter{\romannumeral 1}.}
\end{figure}

Firstly, we investigate the entanglement of two indirectly coupled macroscopic objects. In the experimentally accessible parameter regimes, our scheme realizes the robust entanglement between the movable mirror and atomic ensemble ($N\sim10^7$~\cite{CGDVPTPRA0877,JHJLSCSESPPRL9983}) in the coupled optomechanical system, which is incredible in the macroscopic world. Fig.~2 shows the logarithmic negativity $E_N$ between the movable mirror and atomic ensemble versus the normalized detuning $\Delta/\omega_m$ for different coupling strengths. It can be clearly seen from Fig.~2 that not only the entanglement is increasingly stronger but also the effective detuning is more and more broader with the increase of the coupling strength, which are extremely significant due to the fact that the stronger entanglement and the more broader effective detuning are obtained, the more easily it is realized and observed in experiment. Furthermore, the mirror-atomic ensemble entanglement is present only within a finite interval of values of $\Delta$ around $\Delta \simeq \omega_m$, which is in accordance with Ref.~\cite{DVSGAFHRBPTAGVVAZMAPRL0798}.

\begin{figure}\label{fig3}
\centering
\includegraphics[scale=0.6]{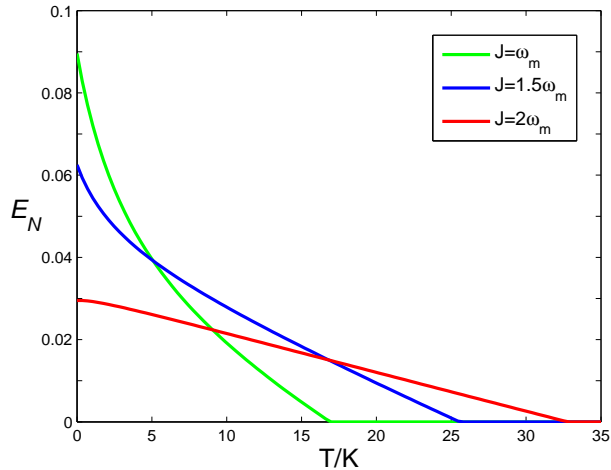}
\caption{(Color online) The logarithmic negativity $E_N$ versus the mirror's temperature $T$ with three different optical coupling strength $J$. Here $\Delta=\omega_m$ and the other parameters are given in Table \uppercase\expandafter{\romannumeral 1}.}
\end{figure}
The robustness of such a mirror-atomic ensemble entanglement with respect to the environmental temperature $T$ of oscillating mirror is shown in Fig.~3. As clearly presented in Fig.~3, due to the environment-induced decoherence, the intensity of the mirror-atomic ensemble entanglement decreases and eventually vanishes with the rise of environmental temperature. With the increase of coupling strength $J$, the critical value of temperature $T_c$ ($T_c$ is defined as $T \geq T_c$, $E_N=0$) increases. When the coupling strength is set $J=2\omega_m$, the critical value of temperature $T_c$ of the mirror-atomic ensemble entanglement persists for 32 K, which is several orders of magnitude larger than the ground state temperature of the mechanical oscillator and is higher than that in Refs.~\cite{DVSGAFHRBPTAGVVAZMAPRL0798,CGDVPTPRA0877}. Therefore, it is easier and more feasible to be realized and observed from the experimental point view.
\begin{figure}\label{fig4}
\centering
\subfigure{\includegraphics[width=2.5in]{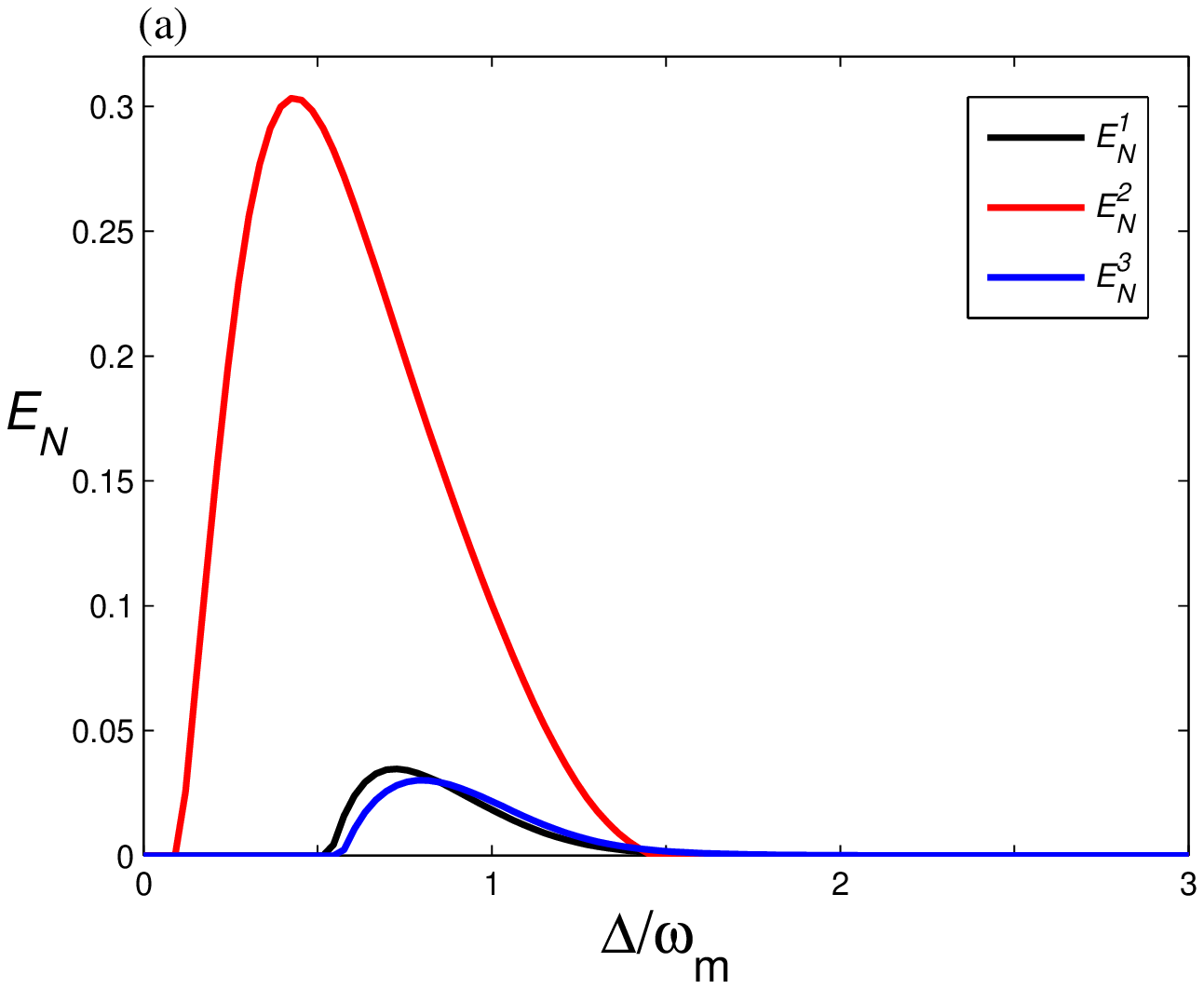}}%
\hspace{0.3in}%
\subfigure{\includegraphics[width=2.5in]{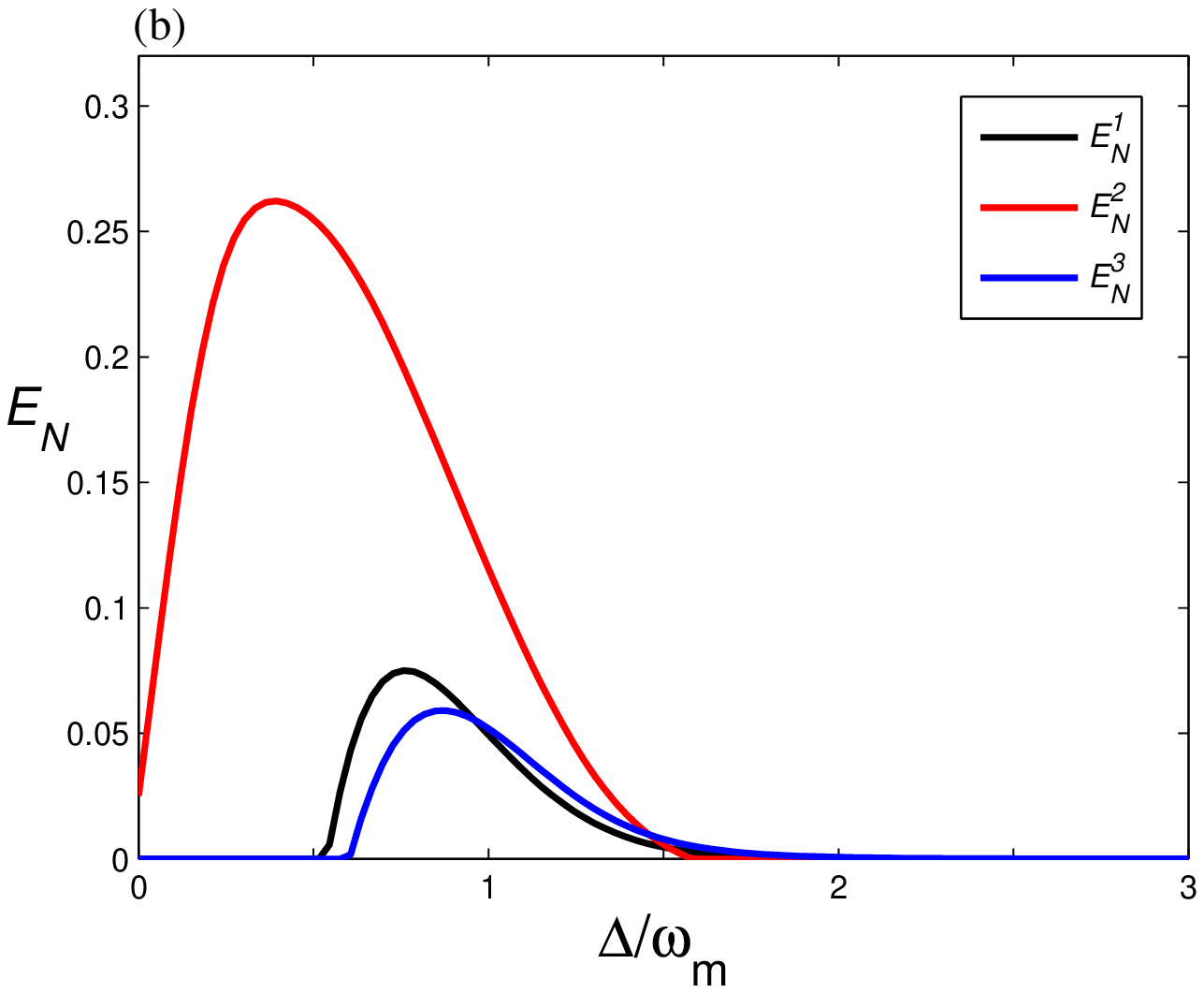}}
\subfigure{\includegraphics[width=2.5in]{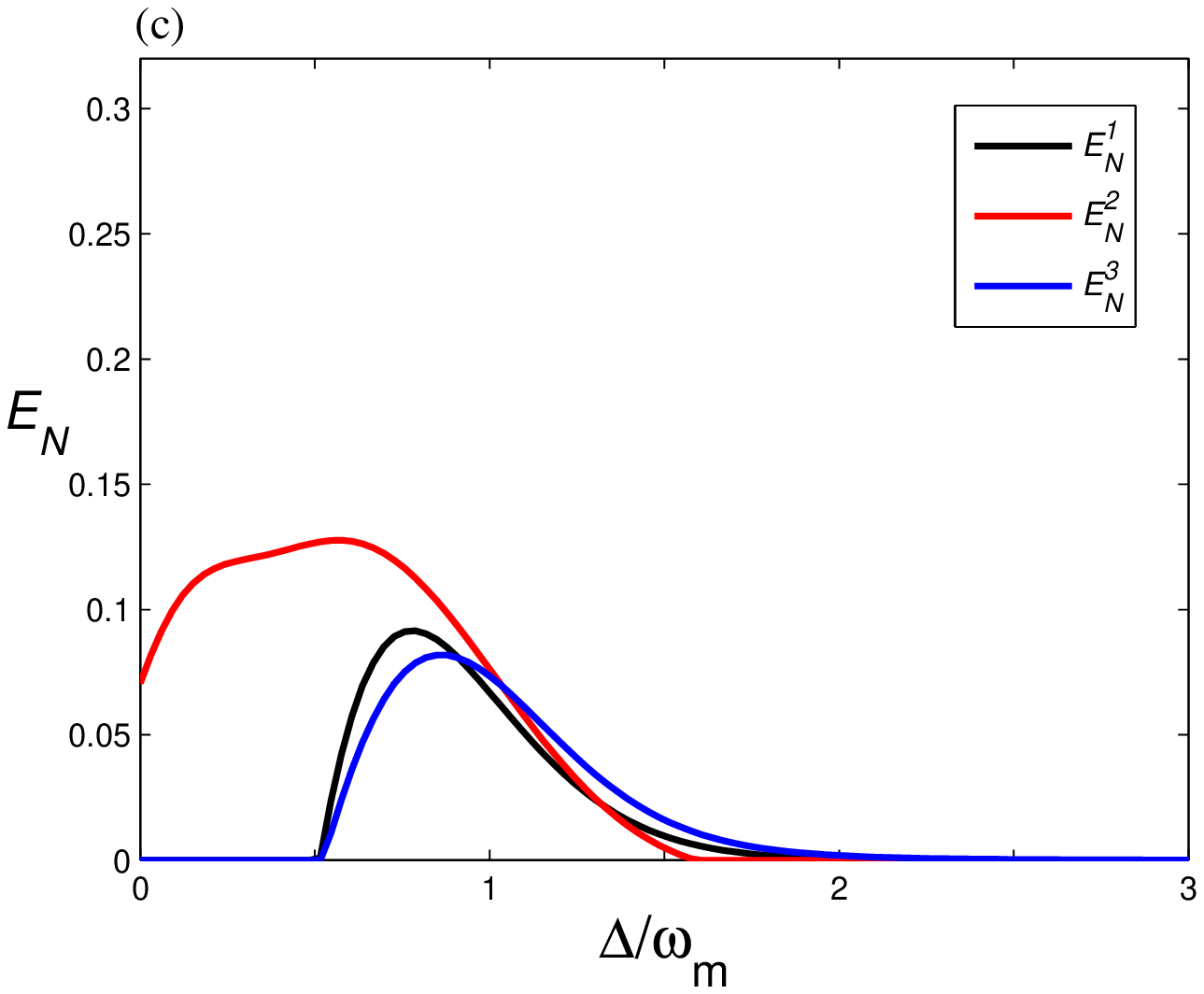}}%
\hspace{0.3in}%
\subfigure{\includegraphics[width=2.5in]{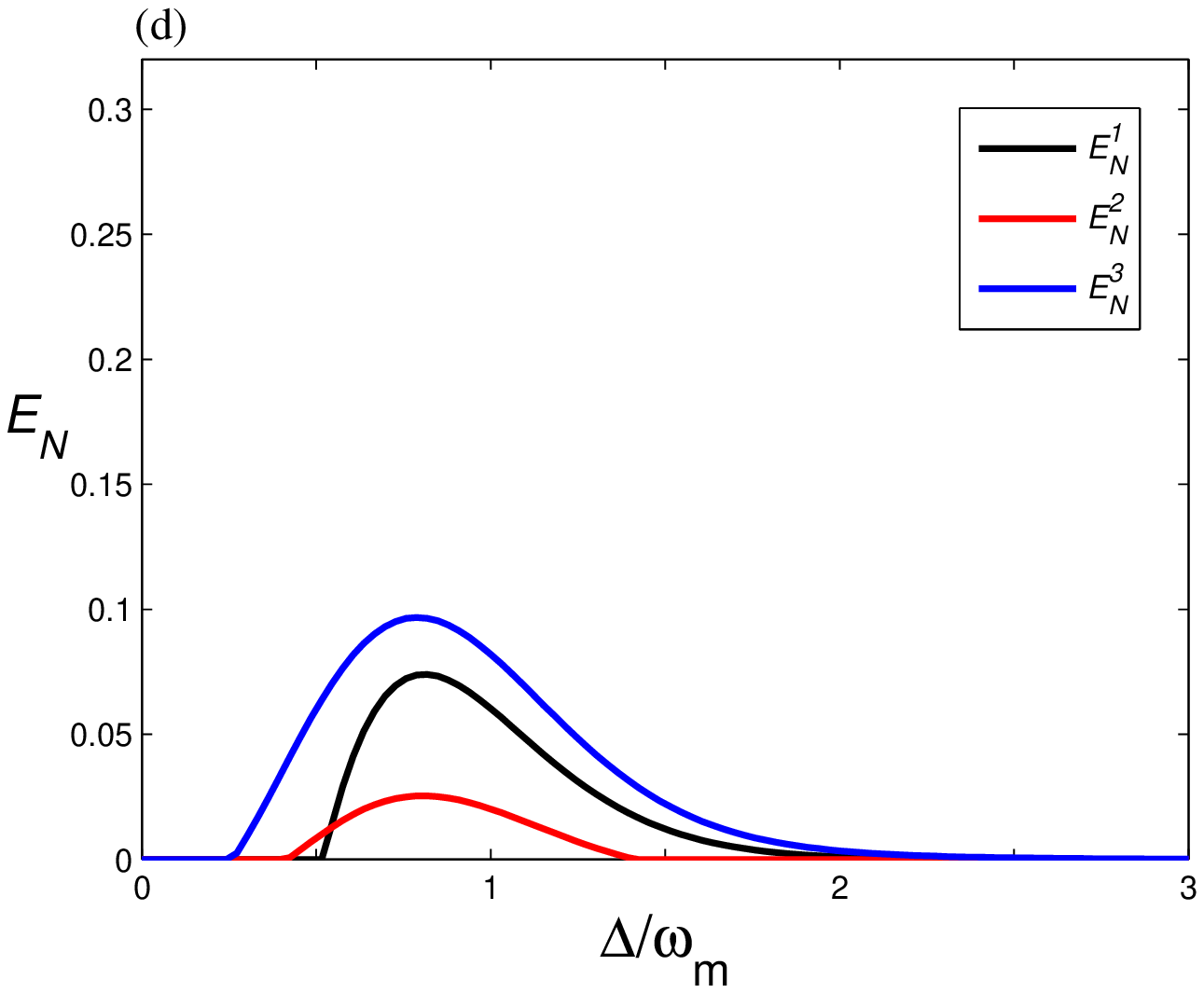}}
\caption{(Color online) The logarithmic negativity $E_N^1$(black curves), $E_N^2$(red curves), and $E_N^3$(blue curves) as a function of the normalized detuning $\Delta/\omega_m$ at a fixed temperature $T$=400 mK and the other parameters are given in Table \uppercase\expandafter{\romannumeral 1}. (a) $J=0.4\omega_m$, (b) $J=0.6\omega_m$, (c) $J=0.8\omega_m$, (d) $J=\omega_m$.}
\end{figure}

Secondly, we investigate the stationary entanglement of the three possible bipartite subsystems in terms of the logarithmic negativity $E_N$. We denote the logarithmic negativities for the cavity 1-mirror, cavity 2-mirror, and atomic ensemble-mirror as $E_N^1$, $E_N^2$, and $E_N^3$, respectively. The bipartite entanglements between the cavity 1-cavity 2, cavity 1-atomic ensemble, and cavity 2-atomic ensemble are so weak that there is no need to consider them. The results on the behavior of the bipartite entanglement are shown in Fig.~4 in which we plot three bipartite logarithmic negativities $E_N^1$(black curves), $E_N^2$(red curves), and $E_N^3$(blue curves) versus the normalized dutuning $\Delta/\omega_m$ at a fixed temperature of $T$=400 mK for four different coupling strengths. It is evident that there is a sort of entanglement transfer among the three bipartite subsystems, i.e., the bipartite entanglements $E_N^1$ and $E_N^3$ increase while the bipartite entanglement $E_N^2$ decreases with the increase of coupling strength $J$. In other words, the enhancement of the entanglements between cavity 1 and mirror and atomic ensemble and mirror at the expense of the entanglement between cavity 2 and mirror. It is remarkable that, therefore, the indirectly coupled entanglement ($E_N^1$ and $E_N^3$) transfers with the directly coupled entanglement ($E_N^2$) each other with the increase of coupling strength $J$. It is worth noting that in the above discussion the effect of the entanglement transfer among the three bipartite subsystems is predominant when the atoms are resonant with the Stokes sideband ($\Delta_a=-\omega_m$). Moreover, we notice that atomic ensemble-mirror entanglement is not present at $\Delta_a=\omega_m$, which is due to the fact that the entanglement is mostly carried by the cavity 1-mirror and cavity 2-mirror.

In above discussion, we assume that the average number of atoms in the excited state is much smaller than the number of total atoms. Here we discuss the limits of validity of the model. The bosonic description of the atomic polarization is valid only when the single-atom excitation probability, $g^2|a_{1s}|^2/(\Delta_a^2+\gamma_a^2)$, is much smaller than 1~\cite{CGDVPTPRA0877}. Furthermore, the linearization of the quantum Langevin equations is valid when the intracavity fields have a large amplitude at the steady state, i.e., $|a_{js}|\gg1$. Therefore, the above two conditions are simultaneously satisfied only when $g^2/(\Delta_a^2+\gamma_a^2)\ll|a_{1s}|^{-2}\ll1$. This means requiring a very weak atom-cavity coupling. But if we consider a relatively small cavity mode volume ($V\simeq10^{-12}$ $\mathrm{m}^3$), in this case, $g$ is not very weak when we consider a standard optical dipole transition. However, the required weak-coupling condition can still be achieved~\cite{CGDVPTPRA0877}.

We now address the experimental issues. The detection of the generated entanglement at the macroscopic level in optomechanical systems is still an experimental challenge. However, for the detection of the entanglement, we have to measure the quadrature correlations~\cite{DFWGJMQO1994} and quantum correlation detection is relatively easy in optomechanical systems. Recently, several promising programs have been proposed in Refs.~\cite{DVSGAFHRBPTAGVVAZMAPRL0798,CGDVPTPRA0877,LMMPPRA1183,MAFSPRA1184}, so we can exploit homodyne measurement experimental techniques to detect quantum correlations so as to detect the indirectly coupled quantum entanglement.

\section{Conclusions}\label{sec5}
In conclusion, we have proposed a scheme to create robust entanglement between a movable mirror and atomic ensemble at the macroscopic level in coupled optomechanical system. With the increase of the coupling strength of the coupled optomechanical system, the stronger entanglement and the broader effective detuning can be obtained, so it is easier and more feasible to realize and observe this sort of novel phenomena in experiment. Utilizing experimentally accessible parameters, the critical temperature of the bipartite continuous variable entanglement in our scheme can approach to 32 K, much higher than that in Refs.~\cite{DVSGAFHRBPTAGVVAZMAPRL0798,CGDVPTPRA0877}. We also investigated the entanglement transfer based on this coupled system. Such a scheme can be used for the realization of quantum memories for continuous variable quantum information processing and quantum-limited displacement measurements.

\begin{center}
{\bf{ACKNOWLEDGMENTS}}
\end{center}

This work was supported by the National Natural Science Foundation of China under
Grant Nos. 11264042, 11465020, 61465013, 11165015, and 11564041.


\begin{thebibliography}{999}
\bibitem{ESPCPSOC3531}E. Schr\"{o}dinger, Proc. Cambridge Philos. Soc. \textbf{31}, 555 (1935).
\bibitem{DBAEAZ2000}D. Bouwmeester, A. Ekert, and A. Zeilinger, {\it The Physics of Quantum Information} (Springer, Berlin, 2000).
\bibitem{MANILC2000}M. A. Nielson and I. L. Chuang, {\it Quantum Computation and Quantum Information} (Camberdge University, 2000).
\bibitem{HASKNJP1113}H. F. Wang, A. D. Zhu, S. Zhang, and K. H. Yeon, New J. Phys. \textbf{13}, 013021 (2011).
\bibitem{WMSSLSZJYLADZHFWSZJOSAB1532}W. M. Sun, S. L. Su, Z. Jin, Y. Liang, A. D. Zhu, H. F. Wang, and S. Zhang, J. Opt. Soc. Am. B \textbf{32}, 9 (2015).
\bibitem{SLSXQSHFWSZPRA1490}S. L. Su, X. Q. Shao, H. F. Wang, and S. Zhang, Phys. Rev. A \textbf{90}, 054302 (2014).
\bibitem{SLSXQSHFWSZSR144}S. L. Su, X. Q. Shao, H. F. Wang, and S. Zhang, Sci. Rep. \textbf{4}, 7566 (2014).
\bibitem{SLSQGHFWSZPRA1592}S. L. Su, Q. Guo, H. F. Wang, and S. Zhang, Phys. Rev. A \textbf{92}, 022328 (2015).
\bibitem{HFWSZPRA0979}H. F. Wang and S. Zhang, Phys. Rev. A \textbf{79}, 042336 (2009).
\bibitem{HFWSZEPJD0953}H. F. Wang and S. Zhang, Eur. Phys. J. D \textbf{53}, 359-363 (2009).
\bibitem{HSAXKOE1119}H. F. Wang, S. Zhang, A. D. Zhu, X. X. Yi, and K. H. Yeon, Opt. Express \textbf{19}, 25433 (2011).

\bibitem{JPPRL9574}J. I. Cirac and P. Zoller, Phys. Rev. Lett. \textbf{74}, 4091 (1995).
\bibitem{DCDN02417}D. Kielpinski, C. Monroe, and D.J. Wineland, Nature (London) \textbf{417}, 709 (2002).
\bibitem{ACWYCKRBEKDLDJWNature14512}A. C. Wilson, Y. Colombe, K. R. Brown, E. Knill, D. Leibfried, and D. J. Wineland, Nature (London) \textbf{512}, 7512 (2014).


\bibitem{DVSGAFHRBPTAGVVAZMAPRL0798}D. Vitali, S. Gigan, A. Ferreira, H. R. B\"{o}hm, P. Tombesi, A. Guerreiro, V.Vedral, A. Zeilinger, and M. Aspelmeyer, Phys. Rev. Lett. \textbf{98}, 030405 (2007)
\bibitem{YHMLZJAP12111}Y. H. Ma and L. Zhou, J. Appl. Phys. \textbf{111}, 103109 (2012).
\bibitem{YCLYWHCWWYFXCPB1322}Y. C. Liu, Y. W. Hu, C. W. Wong, and Y. F. Xiao, Chin. Phys. B \textbf{22}, 114213 (2013).
\bibitem{CGDVPTPRA0877}C. Genes, D. Vitali, and P. Tombesi, Phys. Rev. A \textbf{77}, 050307(R) (2008).
\bibitem{CGAMPTDVPRA0878}C. Genes, A. Mari, P. Tombesi, and D. Vitali, Phys. Rev. A \textbf{78}, 032316 (2008).
\bibitem{CJJLMJEAPOPRA1285}C. Joshi, J. Larson, M. Jonson, E. Andersson, and P. \"{O}hberg, Phys. Rev. A \textbf{85}, 033805 (2012).
\bibitem{UAWMKNGJMPRA1286}U. Akram, W. Munro, K. Nemoto, and G. J. Milburn, Phys. Rev. A \textbf{86}, 042306 (2012).
\bibitem{WGMAAHNMSZPRA1388}W. Ge, M. Al-Amri, H. Nha, and M. S. Zubairy, Phys. Rev. A \textbf{88}, 022338 (2013).
\bibitem{JQLQQWFNPRA1489}J. Q. Liao, Q. Q. Wu, and F. Nori, Phys. Rev. A \textbf{89}, 014302 (2014).
\bibitem{THRZHIPRA1592}T. Huan, R. Zhou, and H. Ian, Phys. Rev. A \textbf{92}, 022301 (2015).
\bibitem{QWYXZMZCPB1524}Q. Wu, Y. Xiao, and Z. M. Zhang, Chin. Phys. B \textbf{24}, 104208 (2015).
\bibitem{ISGHLOPKJVPRL10104}I. S. Grudinin, H. Lee, O. Painter, and K. J. Vahala, Phys. Rev. Lett. \textbf{104}, 083901 (2010).
\bibitem{BPSKOFLFMMGGLLSFFNCMBLYNP1410}B. Peng, S. K. Ozdemir, F. Lei, F. Monifi, M. Gianfreda, G. L. Long, S. Fan, F. Nori, C. M. Bender, and L. Yang, Nat. Phys. \textbf{10}, 394 (2014).
\bibitem{LCXJSHCYJWLJGLGWMXNP148}L. Chang, X. Jiang, S. Hua, C. Yang, J. Wen, L. Jiang, G. Li, G. Wang, and M. Xiao, Nat. Photon. \textbf{8}, 524 (2014).
\bibitem{CKLPRA9551}C. K. Law, Phys. Rev. A \textbf{51}, 2537 (1995).
\bibitem{MPYHAHEPJD997}M. Pinard, Y. Hadjar, A. Heidmann, Eur. Phys. J. D. \textbf{7}, 107 (1999).
\bibitem{SBZPRA1286}S. B. Zheng, Phys. Rev. A \textbf{86}, 013828 (2012).
\bibitem{THHPPR4058}T. Holstein and H. Primakoff, Phys. Rev. \textbf{58}, 1098 (1940).
\bibitem{KHASSESPRMP1082}K. Hammerer, A. S. S{\o}rensen, and E. S. Polzik, Rev. Mod. Phys. \textbf{82}, 1041 (2010).
\bibitem{DFWGJMQO1994}D. F. Walls and G. J. Milburn, {\it Quantum Optics} (Springer, Berlin, 1994).
\bibitem{CWGPZQN2000}C. W. Gardiner and P. Zoller, {\it Quantum Noise} (Springer, Berlin, 2000).
\bibitem{RBMKPRL8146}R. Benguria and M. Kac, Phys. Rev. Lett. \textbf{46}, 1 (1981).
\bibitem{VGDV0163}V. Giovannetti and D. Vitali, Phys. Rev. A \textbf{63}, 023812 (2001).
\bibitem{MATJKFMRMP1486}M. Aspelmeyer, T. J. Kippenberg, and F. Marquardt, Rev. Mod. Phys. \textbf{86}, 1391 (2014).
\bibitem{CFMPSBAHEGST9449}C. Fabre, M. Pinard, S. Bourzeix, A. Heidmann, E. Giacobino, and S. Reynaud, Phys. Rev. A \textbf{49}, 1337 (1994).
\bibitem{EXDKPRA8735}E. X. DeJesus and C. Kaufman, Phys. Rev. A \textbf{35}, 5288 (1987).
\bibitem{PCPVHST1993}P. C. Parks and V. Hahn, {\it Stability Theory} (Prentice Hall, New York, 1993).
\bibitem{MBPMPRL0799}M. Bhattacharya and P. Meystre, Phys. Rev. Lett. \textbf{99}, 153603 (2007).
\bibitem{RSPRL0084}R. Simon, Phys. Rev. Lett. \textbf{84}, 2726 (2000).
\bibitem{MPSVQIC077}M. Plenio and S. Birmani, Quant. Inf. Couput. \textbf{7}, 1 (2007).
\bibitem{GAASFIPRA0470}G. Adesso, A. Serafini, and F. Illuminati, Phys. Rev. A \textbf{70}, 022318 (2004).
\bibitem{DKWMMJADKNDBJPWTMIDBPRL0696}D. Kleckner, W. Marshall, M. J. A. de Dood, K. N. Dinyari, B. J. Pors, W. T. M. Irvine, and D. Bouwmeester, Phys. Rev. Lett. \textbf{96}, 173901 (2006).
\bibitem{SGHRBMPFBGLJBHKCSDBMAAZNature06444}S. Gigan, H. R. Bohm, M. Paternostro, F. Blaser, G. Langer, J. B. Hertzberg, K. C.Schwab, D. B\"{a}uerle, M. Aspelmeyer, and A. Zeilinger, Nature (London) \textbf{444}, 67 (2006).
\bibitem{OAPFCTBMPAHNature06444}O. Arcizet, P.-F. Cohadon, T. Briant, M. Pinard, and A. Heidmann, Nature (London) \textbf{444}, 71 (2006).
\bibitem{TYWZLZ1461}T. Yousif, W. Zhou, and L. Zhou, J. Mod. Opt. \textbf{61}, 14 (2014).
\bibitem{JHJLSCSESPPRL9983}J. Hald, J. L. S{\o}rensen, C. Schori, and E. S. Polzik, Phys. Rev. Lett. \textbf{83}, 1319 (1999).
\bibitem{LMMPPRA1183}L. Mazzola and M. Paternostro, Phys. Rev. A \textbf{83}, 062335 (2011).
\bibitem{MAFSPRA1184}M. Asjad and F. Saif, Phys. Rev. A \textbf{84}, 033606 (2011).
\end{thebibliography}
\end{document}